# YARKOVSKY EFFECT IN GENERALIZED PHOTOGRAVITATIONAL 3-BODIES PROBLEM

**Sergey V. Ershkov**,

Institute for Time Nature Explorations,

M.V. Lomonosov's Moscow State University,

Leninskie gory, 1-12, Moscow 119991, Russia

e-mail: sergej-ershkov@yandex.ru

**Abstract:** Here is presented a generalization of photogravitational restricted 3-bodies problem to the case of influence of Yarkovsky effect, which is known as reason of additional infinitesimal acceleration of a small bodies in the space (due to anisotropic re-emission of absorbed energy from the sun, other stellar sources).

Asteroid is supposed to move under the influence of gravitational forces from 2 massive bodies (*which are rotating around their common centre of masses on Kepler's trajectories*), as well under the influence of pressure of light from both the primaries.

Analyzing the ODE system of motion, we explore the existense of equilibrium points for a small body (asteroid) in the case when the 2-nd primary is *non-oblate* spheroid.

In such a case, it is proved the existence of maximally *256* different *non-planar* libration points in generalized photogravitational restricted 3-bodies problem when we take into consideration even a small *Yarkovsky effect*.

# 1. Introduction.

The Yarkovsky effect is a force acting on a rotating body in space caused by the anisotropic emission of thermal photons, which carry momentum (Radzievskii 1954). It is usually considered in relation to meteoroids or small asteroids (*about 10 cm to 10 km in diameter*), as its influence is most significant for these bodies. Such a force is produced by the way an asteroid absorbs energy from the sun and re-radiates it into space as heat by anisotropic way.

In fact, there exists a disbalance of momentum when asteroid at first absorbs the light, radiating from the sun, but then asteroid re-radiates the heat. Such a disbalance is caused by the rotating of asteroid during period of warming as well as it is caused by the anisotropic cooling of surface & inner layers; the processes above depend on anisotropic heat transfer in the inner layers of asteroid.

During thousands of years such a disbalance forms a negligible, but important additional acceleration for a small bodies, so-called Yarkovsky effect. Thus, Yarkovsky effect is small but very important effect in celestial mechanics as well as in calculating of a proper orbits of asteroids & other small bodies.

Besides, Yarkovsky effect *is not predictable* (*it could be only observed & measured by astronomical methods*); the main reason is unpredictable character of the rotating of small bodies (Rubincam 2000), even in the case when there is no any collision between them.

If regime of the rotating of asteroid is changing, we could observe a generalization of Yarkovsky effect, i.e. the Yarkovsky–O'Keefe–Radzievskii–Paddack effect or YORP effect (Rubincam 2000).

## 2. Equations of motion.

Let us consider the system of ordinary differential equations for photogravitational restricted 3-bodies problem, at given initial conditions (Radzievskii 1950; Shankaran *et al.* 2011).

In according with (Shankaran *et al.* 2011), we consider three bodies of masses $m_1$, $m_2$ and $m$ such that $m_1 > m_2$ and $m$ is an infinitesimal mass. The two primaries $m_1$ and $m_2$ are sources of radiation; $q_1$ and $q_2$ are factors characterizing the radiation effects of the two primaries respectively, $\{q_1, q_2\} \in (-\infty, 1]$. We assume that $m_2$ is an *oblate* spheroid. The effect of *oblateness* is denoted by the factor $A_2$. Let $r_i$ ($i = 1, 2$) be the distances between the centre of mass of the bodies $m_1$ and $m_2$ and the centre of mass of body $m$ (Shankaran *et al.* 2011).

Now, the unit of mass is chosen so that the sum of the masses of finite bodies is equal to 1. We suppose that $m_1 = 1 - \mu$ and $m_2 = \mu$, where $\mu$ is the ratio of the mass of the smaller primary to the total mass of the primaries and $0 \leq \mu \leq 1/2$. The unit of distance is taken as the distance between the primaries. The unit of time is chosen so that the gravitational constant is equal to 1 (Shankaran *et al.* 2011).

The three dimensional restricted 3-bodies problem, with an *oblate* primary $m_2$ and both primaries radiating, could be presented in barycentric rotating co-ordinate system by the equations of motion below (Shankaran *et al.* 2011; Douskos *et al.* 2006):

$$\ddot{x} - 2n\dot{y} = \frac{\partial \Omega}{\partial x},$$

$$\ddot{y} + 2n\dot{x} = \frac{\partial \Omega}{\partial y}, \qquad (2.1)$$

$$\ddot{z} = \frac{\partial \Omega}{\partial z},$$

$$\Omega = \frac{n^2}{2}\left(x^2+y^2\right)+\frac{q_1(1-\mu)}{r_1}+\frac{q_2\mu}{r_2}\left[1+\frac{A_2}{2r_2^2}\cdot\left(1-\frac{3z^2}{r_2^2}\right)\right], \quad (2.2)$$

- where

$$n^2 = 1 + \frac{3}{2}A_2,$$

- is the angular velocity of the rotating coordinate system and $A_2$ - is the *oblateness* coefficient. Here

$$A_2 = \frac{AE^2 - AP^2}{5R^2},$$

- where *AE* is the equatorial radius, *AP* is the polar radius and *R* is the distance between primaries. Besides, we should note that

$$r_1^2 = (x+\mu)^2 + y^2 + z^2,$$

$$r_2^2 = (x-1+\mu)^2 + y^2 + z^2,$$

- are the distances of infinitesimal mass from the primaries.

We neglect the relativistic Poynting-Robertson effect which may be treated as a perturbation for cosmic dust (*or for small particles, less than 1 cm in diameter*), see Chernikov (Chernikov 1970; Kushvah *et al.* 2007), as well as we neglect the effect of variable masses of 3-bodies (Singh *et al.* 2010).

## 3. Modified equations of motion (Yarkovsky effect).

Modified equations of motion (2.1) for the generalized three dimensional restricted 3-bodies problem, with an *oblate* primary $m_2$, both primaries radiating, and the infinitesimal mass *m* under the influence of Yarkovsky effect, should be presented in barycentric rotating co-ordinate system in the form below:

$$\ddot{x} - 2n\dot{y} = \frac{\partial \Omega}{\partial x} + Y_x(t),$$

$$\ddot{y} + 2n\dot{x} = \frac{\partial \Omega}{\partial y} + Y_y(t), \qquad (3.1)$$

$$\ddot{z} = \frac{\partial \Omega}{\partial z} + Y_z(t),$$

- where $Y_x(t)$, $Y_y(t)$, $Y_z(t)$ – are the projecting of Yarkovsky effect acceleration *Y (t)* on the appropriate axis *Ox, Oy, Oz*.

## 4. Location of Equilibrium points.

The location of equilibrium points for system (3.1) in general is given by conditions:

$$\ddot{x} = \ddot{y} = \ddot{z} = \dot{x} = \dot{y} = \dot{z} = 0,$$

(4.1)

$$\frac{\partial \Omega}{\partial x} = -Y_x(t), \quad \frac{\partial \Omega}{\partial y} = -Y_y(t), \quad \frac{\partial \Omega}{\partial z} = -Y_z(t).$$

Let us consider the case when the effect of *oblateness* is absent, $A_2 = 0$ ($\Rightarrow n = 1$), see the appropriate expression:

$$n^2 = 1 + \frac{3}{2}A_2$$

It means a significant simplifying of expression (2.2) in the system of equalities (4.1):

$$-Y_x = x - \frac{q_1(1-\mu)\cdot(x+\mu)}{\left((x+\mu)^2 + y^2 + z^2\right)^{\frac{3}{2}}} - \frac{q_2\mu\cdot(x-1+\mu)}{\left((x-1+\mu)^2 + y^2 + z^2\right)^{\frac{3}{2}}},$$

$$-Y_y = y - \frac{q_1(1-\mu)\cdot y}{\left((x+\mu)^2 + y^2 + z^2\right)^{\frac{3}{2}}} - \frac{q_2\mu\cdot y}{\left((x-1+\mu)^2 + y^2 + z^2\right)^{\frac{3}{2}}},$$

$$-Y_z = -\frac{q_1(1-\mu)\cdot z}{\left((x+\mu)^2 + y^2 + z^2\right)^{\frac{3}{2}}} - \frac{q_2\mu\cdot z}{\left((x-1+\mu)^2 + y^2 + z^2\right)^{\frac{3}{2}}}.$$

Besides, we assume all equations (4.1) to be *a united system* of algebraic equations. That's why we substitute an expression for $z$ from 3-rd equation above to the 2-nd & the 1-st equation:

$$z \cdot Y_x = z \cdot \left(-x + \frac{q_1(1-\mu)}{\left((x+\mu)^2 + y^2 + z^2\right)^{\frac{3}{2}}}\right) + Y_z \cdot (x-1+\mu),$$

$$-z \cdot Y_y = y \cdot (z - Y_z), \quad \Rightarrow \quad Y_z \cdot y = z \cdot (y + Y_y), \qquad (4.2)$$

$$Y_z = z \cdot \left(\frac{q_1(1-\mu)}{\left((x+\mu)^2 + y^2 + z^2\right)^{\frac{3}{2}}} + \frac{q_2\mu}{\left((x-1+\mu)^2 + y^2 + z^2\right)^{\frac{3}{2}}}\right).$$

Moreover, we obtain from the 3-d equation of system (4.2) that *planar* equilibrium points exist only if $\{Y_z = 0,\ z = 0\}$ simultaneously. But the case $Y_z = 0$ – is very rare, specific condition for asteroid, which has unpredictable character of the regime of rotating during a flight through the space (Rubincam 2000); the same is obtained for the case $y = 0$.

Therefore we will consider only *non-planar* equilibrium points $z, y \neq 0$. So, we obtain from the 1-st & 3-d equations of system (4.2):

$$\frac{q_1(1-\mu)}{\left((x+\mu)^2 + y^2 + z^2\right)^{\frac{3}{2}}} = Y_x + x - Y_z \cdot \frac{(x-1+\mu)}{z},$$

$$z = Y_z \cdot \frac{y}{(y + Y_y)}, \quad y \neq -Y_y, \qquad (4.3)$$

$$\frac{q_2 \mu}{\left((x-1+\mu)^2 + y^2 + z^2\right)^{\frac{3}{2}}} = -Y_x - x + Y_z \cdot \frac{(x+\mu)}{z}.$$

Hence, we finally obtain the system of algebraic equations for meanings of $\{x, y\}$, which determine the location of equilibrium points (4.1):

$$\begin{cases} \dfrac{q_1(1-\mu)}{\left((x+\mu)^2 + y^2 + z^2\right)^{\frac{3}{2}}} = Y_x + x - Y_z \cdot \dfrac{(x-1+\mu)}{z}, \\ \\ \dfrac{q_2 \mu}{\left((x-1+\mu)^2 + y^2 + z^2\right)^{\frac{3}{2}}} = -Y_x - x + Y_z \cdot \dfrac{(x+\mu)}{z}, \end{cases} \qquad (4.4)$$

- where

$$z = Y_z \cdot \frac{y}{(y + Y_y)}, \quad y \neq -Y_y \ .$$

The last system (4.4) could be presented as below:

$$\begin{cases} \dfrac{q_1^2 (1-\mu)^2 \cdot (y+Y_y)^6 \cdot y^2}{\left( \left((x+\mu)^2 + y^2\right) \cdot (y+Y_y)^2 + Y_z^2 \cdot y^2 \right)^3} = \left( y \cdot (Y_x + x) - (x - 1 + \mu) \cdot (y + Y_y) \right)^2 , \\ \\ \dfrac{q_2^2 \mu^2 \cdot (y+Y_y)^6 \cdot y^2}{\left( \left((x-1+\mu)^2 + y^2\right) \cdot (y+Y_y)^2 + Y_z^2 \cdot y^2 \right)^3} = \left( -y \cdot (Y_x + x) + (x + \mu) \cdot (y + Y_y) \right)^2 , \end{cases}$$

- where the maximal polynomial order of equations is equal to 16 x 16 = 256: indeed, the order of 1-st polynomial equation is equal to 16 (*in regard to variables x,y*); the order of 2-nd polynomial equation is also equal to 16 (*in regard to x,y*).

So, (4.4) is the polynomial system of equations of *256-th* order which has maximally *256* different roots, we should especially note that each of them strongly depends on various parameters { $\mu$, $q_1$, $q_2$ ; $Y_x$, $Y_y$, $Y_z$ }. Such a system of polynomial equations could be solved only by numerical methods (*in general, it is valid for polynomial equation of order > 5*).

Besides, analysing the equations of system (4.2), we should note that a case of *Yarkovsky effect is negligible* determines the existence of *quasi-planar* equilibrium points in which conditions {$Y_z \to 0$, $z \to 0$} are valid *simultaneously*.

To give some estimation or numerical results, we should take into consideration *the negligible character* of Yarkovsky effect $\{Y_x, Y_y, Y_z\} \to 0$ in the last system of equation of 256-th order.

Such a simplification let us obtain the result below ($y \neq 0$, see (4.2)):

$$\begin{cases} \dfrac{q_1^2}{\left((x+\mu)^2+y^2\right)^3} = 1 + \dfrac{2Y_x}{(1-\mu)} + \left(1 - \dfrac{x}{(1-\mu)}\right) \cdot \dfrac{2Y_y}{y}, \\ \\ \dfrac{q_2^2}{\left((x-1+\mu)^2+y^2\right)^3} = 1 - \dfrac{2Y_x}{\mu} + \dfrac{(x+\mu)}{\mu} \cdot \dfrac{2Y_y}{y}, \end{cases} \qquad (4.5)$$

If we substitute the appropriate meanings of coordinates *x, y* for triangular libration points *L4* and *L5* in (4.5) when Yarkovsky effect equals to zero, we will obtain that all the equalities are valid in terms of generalized photogravitational restricted 3-bodies problem (Xuetang *et al.* 1993).

Each of equations of system (4.5) has 7-th order (*in regard to variables x,y*), so (4.5) is the polynomial system of equations of *49-th* order which has maximally *49* different roots. That's why let us make the next step for simplifying of the system (4.5):

$$\begin{cases} \dfrac{q_1^{\frac{2}{3}}}{\left((x+\mu)^2+y^2\right)} = 1 + \dfrac{2Y_x}{3(1-\mu)} + \left(1 - \dfrac{x}{(1-\mu)}\right) \cdot \dfrac{2Y_y}{3y}, \\ \\ \dfrac{q_2^{\frac{2}{3}}}{\left((x-1+\mu)^2+y^2\right)} = 1 - \dfrac{2Y_x}{3\mu} + \dfrac{(x+\mu)}{\mu} \cdot \dfrac{2Y_y}{3y}, \end{cases} \qquad (4.6)$$

Each of equations of system (4.6) has *3-d* order (*in regard to variables x,y*), so (4.6) is the polynomial system of equations of *6-th* order which has maximally *6* different roots. Such a system of polynomial equations could be also solved only by numerical methods (*it is valid for polynomial equation of order > 5*).

Let us present the solution which differ from the libration points $L_4$ and $L_5$ (*due to Yarkovsky effect*) as below:

$$x = x_0 + \Delta x, \quad y = y_0 + \Delta y,$$

- where $x_0$, $y_0$ – the appropriate meanings of coordinates of the triangular libration points $L_4$ and $L_5$ in generalized photogravitational restricted 3-bodies problem, when $Y_x = Y_y = Y_z = 0$ (Xuetang *et al.* 1993). So, from (4.6) we obtain ($\Delta x, \Delta y \to 0$):

$$\begin{cases} \Delta x \cdot 1 = \frac{q_1^{\frac{2}{3}}}{2} \cdot \left( -\frac{2Y_x}{3(1-\mu)} - \left(1 - \frac{x_0}{(1-\mu)}\right) \cdot \frac{2Y_y}{3y_0} \right) - \frac{q_2^{\frac{2}{3}}}{2} \cdot \left( \frac{2Y_x}{3\mu} - \frac{(x_0+\mu)}{\mu} \cdot \frac{2Y_y}{3y_0} \right), \\ \\ \Delta y \cdot y_0 = \frac{q_1^{\frac{2}{3}}}{2} \cdot \left( -\frac{2Y_x}{3(1-\mu)} - \left(1 - \frac{x_0}{(1-\mu)}\right) \cdot \frac{2Y_y}{3y_0} \right) - \Delta x \cdot (x_0 + \mu). \end{cases}$$

The strongest simplifying of system (4.4) is possible when *Yarkovsky effect is zero*, $Y_x = Y_y = Y_z = 0$. In such a case, it has been proved the existence of maximally *9* different equilibrium points $\{L_1, ..., L_9\}$ in photogravitational restricted 3-bodies problem (Xuetang *et al.* 1993).

# 5. Conclusion.

It has been proved the existence of maximally *256* different *non-planar* equilibrium points $z, y \neq 0$ in generalized photogravitational restricted 3-bodies problem when we take into consideration even a small *Yarkovsky effect* in the case the 2-nd primary is *non-oblate* spheroid. This result is different both from classical restricted 3-bodies problem and generalized photogravitational restricted 3-bodies problem.

Stability of such a points is an open problem in celestial mechanics for the case of non-zero *Yarkovsky effect* (Radzievskii 1950).

This model may be applied to examine the dynamic behaviour of small rotating objects such as meteoroids or small asteroids (*about 10 cm to 10 km in diameter*).

For the meteoroids less than 10 cm in diameter we should additionally take into consideration the relativistic Poynting-Robertson effect which may be treated as a perturbation for cosmic dust, see Chernikov (Chernikov 1970; Kushvah *et al.* 2007).

Yarkovsky effect does not make any significant influence in regard to the meteoroids more than 10 km in diameter (Radzievskii 1954).

# Acknowledgements

I am thankful to CNews Russia project (*Science & Technology Forum, branch "Gravitation"*) - for valuable discussions in preparing this manuscript. Especially I am thankful to Dr. P.Fedotov, Col. L.Vladimirov, Dr. A.Kulikov for valuable suggestions in preliminary discussions of this manuscript.